\documentclass{revtex4}
\usepackage{epsfig}
\usepackage{amssymb}
\usepackage{amsmath}
\usepackage{amsfonts}
\usepackage{graphicx}
\usepackage{mathrsfs}
\usepackage[dvips]{color}
\usepackage{multirow}

\newcommand{\R}{\mathbb{R}}
\newcommand{\C}{\mathbb{C}}

\newcommand{\fa}{\mathfrak{a}}
\newcommand{\fb}{\mathfrak{b}}

\newcommand{\fn}{{\mathfrak{n}}}

\newcommand{\fz}{\mathfrak{z}}

\newcommand{\bI}{\mathbf{I}}

\newcommand{\bM}{\mathbf{M}}

\newcommand{\cK}{\mathcal{K}}

\newcommand{\cP}{\mathcal{P}}

\newcommand{\cT}{\mathcal{T}}

\newcommand{\be}{\begin{equation}}
\newcommand{\ee}{\end{equation}}
\newcommand{\bea}{\begin{eqnarray}}
\newcommand{\eea}{\end{eqnarray}}
\newcommand{\nn}{\nonumber}

\newcommand{\ed}{\end{document}}

\newcommand{\bi}{\begin{itemize}}
\newcommand{\ei}{\end{itemize}}

\newcommand{\bce}{\begin{center}}
\newcommand{\ece}{\end{center}}

\newcommand{\sE}{\mathscr{E}}

\newcommand{\sH}{\mathscr{H}}

\newcommand{\sT}{\mathscr{T}}
\newcommand{\sU}{\mathscr{U}}

\begin{document}

\title{Perturbative Unidirectional Invisibility}

\author{Ali~Mostafazadeh}
\address{Departments of Mathematics and Physics, Ko\c{c}
University,\\ Sar{\i}yer 34450, Istanbul, Turkey\\
amostafazadeh@ku.edu.tr}

\begin{abstract}
We outline a general perturbative method of evaluating scattering features of finite-range complex potentials and use it to examine complex perturbations of a rectangular barrier potential. In optics, these correspond to modulated refractive index profiles of the form
$\fn(x)=n_0+f(x)$, where $n_0$ is real, $f(x)$ is complex-valued, and $|f(x)|\ll1\leq n_0$. We give a comprehensive description of the phenomenon of unidirectional invisibility for such media, proving five general theorems on its realization in $\cP\cT$-symmetric and non-$\cP\cT$-symmetric material. In particular, we establish the impossibility of unidirectional invisibility for $\cP\cT$-symmetric samples whose refractive index has a constant real part and show how a simple scaling transformation of a unidirectionally invisible  $\cP\cT$-symmetric index profile with $n_0=1$ may be used to generate a hierarchy of unidirectionally invisible $\cP\cT$-symmetric index profiles with $n_0>1$. The results pertaining unidirectional invisibility for $n_0>1$ open up the way for the experimental studies of this phenomenon in a variety of active material. As an application of our general results, we show that a medium with $\fn(x)=n_0+\zeta e^{iK x}$, $\zeta$ and $K$ real, and $|\zeta|\ll 1$ can support unidirectional invisibility only for $n_0=1$. We then construct unidirectionally invisible index profiles of the form $\fn(x)=n_0+\sum_\ell \fz_\ell e^{iK_\ell x}$, with $\fz_\ell$ complex, $K_\ell$ real, $|\fz_\ell|\ll 1$, and $n_0>1$.
\\

\noindent {Pacs numbers: 42.25.Bs, 03.65.Nk, 02.30.Zz}\\

\noindent Keywords: Perturbative scattering, optical potential, unidirectional invisibility, $\cP\cT$-symmetry

\medskip

%\noindent {Pacs numbers: 03.65.Nk}
\end{abstract}

\maketitle

\section{Introduction}

Time-independent scattering theory is a well-established discipline with numerous applications in different areas of physics and engineering. The standard textbook treatments of this theory are usually confined to real scattering potentials, but the generalization to complex potentials does not cause any sever difficulties. Recent years have witnessed a growing interest in the study of complex scattering potentials, because unlike real potentials they are capable of supporting interesting phenomena such as spectral singularities \cite{prl-2009,ss,pra-2011a,jpa-2012} and unidirectional invisibility \cite{invisible-0,lin,invisible-1,pra-2013a}. Spectral singularities correspond to scattering states which behave like zero-width resonances \cite{prl-2009,prl-2013}. They provide a
mathematical description of lasing at the threshold gain \cite{pra-2011a} and anti-lasing \cite{longhi-2010}. Unidirectional invisibility is the property of having perfect transmission and unidirectional reflection. The possibility of realizing it has attracted a lot of attention, because it provides a tool for constructing certain one-way optical devices \cite{feng-yin}. Another remarkable property of unidirectionally invisible potentials is that they serve as the building blocks for constructing potentials with given scattering properties at a given wavenumber \cite{pra-2014bc}. These observations provide ample motivation for a systematic study of the problem of characterizing scattering potentials displaying unidirectional invisibility. The purpose of this article is to propose a solution to this problem which can be conveniently employed in a wide range of easily-realizable optical setups.

Consider the Helmholtz equation
    \be
    \sE''(x)+k^2 \fn(x)^{2} \sE(x)=0,
    \label{HE}
    \ee
which describes the propagation of time-harmonic electromagnetic waves interacting with an infinite planar slab of optically active material. Here the electric field is given by $\vec E(x,t)=e^{-i k c t}\sE(x)\hat e_y$, $\hat e_y$ is the unit vector along the positive $y$-axis in some Cartesian coordinate system, $\{(x,y,z)\}$, $k$ is the wavenumber, $c$ is the speed of light in vacuum, and $\fn(x)$ is the refractive index of the medium. For a slab consisting of material obtained by doping a homogeneous host medium with index of refraction $\fn_{\!0}$, the latter has the form
    \be
    \fn(x)=\left\{\begin{array}{cc}
    \fn_{\!0}+\nu(x)+i\kappa(x) &{\rm for}~x\in[0,L],\\[6pt]
    1&{\rm otherwise},\end{array}\right.
    \label{n2=x}
    \ee
where $\nu$ and $\kappa$ are piecewise continuous real-valued functions with support $[0,L]$, and $L$ is the thickness of the slab. Let $n_0$ and $n'_0$ respectively denote the real and imaginary part of $\fn_0$, so that
    \be
    \fn_0=n_0+i n'_0.
    \label{n0=}
    \ee
Then the regions in the optically active part of the space, i.e., $[0,L]$, in which $n'_0+\kappa(x)$ takes negative (respectively positive) values display gain (respectively loss) properties. For a non-exotic active material, $n_0\geq 1 $, and $|n'_0|$, $|\nu(x)|$, and $|\kappa(x)|$ are at least three orders of magnitude smaller than $n_0$;
    \be
    |n'_0|+|\nu(x)|+|\kappa(x)|\ll 1\leq n_0.
    \label{condi}
    \ee

We can identify (\ref{HE}) with the time-independent Schr\"odinger equation,
    \be
    -\psi''(x)+v(x)\psi(x)=k^2\psi(x),
    \label{SE}
    \ee
for the complex barrier potential:
    \be
    v(x)=k^2[1-\fn(x)^2]\approx \left\{\begin{array}{ccc}
    v_0+v_1(x) &{\rm for}& x\in[0,L],\\
    0&{\rm for}& x\notin[0,L],\end{array}\right.
    \label{v=21}
    \ee
where we have employed (\ref{condi}) and introduced
    \begin{align}
    &v_0:=k^2(1-\fn_{\!0}^2), &&
    v_1(x):=-2k^2\fn_{\!0}[\nu(x)+i\kappa(x)].
    \label{v=21-def}
    \end{align}

In Ref.~\cite{invisible-0,lin,invisible-1}, the authors note that for
    \begin{align}
    &v_0(x)=0, &&v_1(x)=\fz\,e^{2\pi i x/L},
    \label{exp-potential}
    \end{align}
$\fz\in\R^+$, $\fz\ll 1$, and $k=\pi/L$, the potential (\ref{v=21}) displays unidirectional invisibility. This corresponds to a slab with $\fn_0=n_0=1$ for which we can employ the perturbative treatment of the scattering phenomenon as outlined in Ref.~\cite{pra-2014a}. It turns out that this potential violates the condition of perfect transmission, if one goes beyond the first order perturbations theory, i.e., it displays perturbative unidirectional invisibility \cite{pra-2014a}. An experimental verification of this behavior requires modulating both the real and imaginary parts of the refractive index $\fn(x)$. The latter is usually performed by pumping the active media to maintain the desired gain/loss profile for the slab. Manufacturing samples with particular shape for the real part of the refractive index requires other means \cite{Hang}.

To the best of our knowledge, except for the results obtained for bilayer slabs in \cite{pra-2013a}, the recent theoretical developments \cite{invisible-0,lin,invisible-1,pra-2014a} in the study of the unidirectional invisibility are obtained under the assumption that $n_0=1$. This imposes severe limitations on the direct experimental manifestations of this phenomenon, because it restricts the choice of the host medium and subsequently puts strong bounds on the attainable values of the imaginary part of $\fn(x)$. In this article, we avoid these limitation by offering a generalization of the perturbative approximation scheme developed in \cite{pra-2014a} to situations where $n_0>1$. This allows for a comprehensive study of the subject and reveals a number of remarkable properties of the finite-range $\cP\cT$-symmetric and non-$\cP\cT$-symmetric optical potentials supporting perturbative unidirectional invisibility.

In the remainder of this section we survey some of the basic properties of the transfer matrices, give a precise definition of unidirectional invisibility, and provide a brief review of a recently proposed dynamical formulation of time-independent scattering theory which serves as the basic theoretical framework for the developments we report in this article.

For a general real or complex scattering potential $v(x)$ and a real wavenumber $k$, the solutions of the Schr\"odinger equation (\ref{SE}) have the following asymptotic form \footnote{By a scattering potential we mean a function $v:\R\to\C$ fulfilling the decay condition $\int_{-\infty}^\infty dx\, (1+|x|)|v(x)|<\infty$, so that the Schr\"odinger equation (\ref{SE}) admits scattering (Jost) solutions \cite{kemp}.}
    \be
    \psi(x)\to A_\pm e^{ikx}+B_\pm e^{-ikx}~~{\rm for}~~x\to\pm\infty.
    \label{psi=}
    \ee
The $2\times 2$ matrix $\bM$ satisfying
    \be
    \bM\left[\begin{array}{c} A_-\\ B_-\end{array}\right]=\left[\begin{array}{c} A_+\\ B_+\end{array}\right]
    \label{defn}
    \ee
is called the transfer matrix of $v(x)$. Its entries $M_{ij}$ are related to the reflection and transmission amplitudes, $R^{l/r}$ and $T$, of the potential $v(x)$ according to \cite{prl-2009}:
    \begin{align}
	&M_{11}=T-\frac{R^l R^r}{T}, &&
	M_{12}=\frac{R^r}{T}, && M_{21}=-\frac{R^l}{T}, && M_{22}=\frac{1}{T}.
	\label{M-RT}
	\end{align}
As a simple consequence of (\ref{M-RT}) we also recover the well-known identity $\det\bM=1$, \cite{jpa-2009,sanchez}.

The transfer matrix and the reflection and transmission amplitudes are functions of the wavenumber. Suppose that we fix a particular value $k_\star$ for the latter. Then, by definition, $v(x)$ is said be unidirectionally reflectionless from the left (respectively right) or simply left-reflectionless (respectively  right-reflectionless)  for $k=k_\star$, if $R^l(k_\star)=M_{21}(k_\star)=0$ (resp.\ $R^r(k_\star)=M_{12}(k_\star)=0$). It is called unidirectionally invisible from the left (respectively  right) or simply left-invisible (respectively right-invisible)  at the wavenumber $k_\star$, if in addition we have $T(k_\star)=M_{22}(k_\star)=M_{11}(k_\star)=1$.

In Refs.~\cite{ap-2014,pra-2014a}, we propose a dynamical formulation of the one-dimensional potential scattering that identifies the transfer matrix $\bM$ of a complex scattering potential $v(x)$ with the asymptotic time-evolution operator for a non-unitary two-level quantum system. More specifically, $\bM=\sU(\infty,-\infty)$, where
	\[\sU(\tau,\tau_0):=\sT\exp\left\{-i\int_{\tau_0}^\tau d\tau' \sH(\tau')\right\}\]
is the time-evolution operator for the non-Hermitian Hamiltonian operator:
    \be
    \sH(\tau):=w(\tau)\,\cK(\tau),
	\label{int-H}
    \ee
$w(\tau)$ and $\cK(\tau)$ are given by
    \begin{align}
    &w(\tau):=\frac{v(\tau/k)}{2k^2}=\frac{1-\fn(\tau/k)^2}{2},
     \label{w}\\
     & \cK(\tau):=\left[\begin{array}{cc}
	1 & e^{-2i\tau}\\
	-e^{2i \tau} & -1\end{array}\right],
	\label{K=}
    \end{align}
and $\sT$ denotes the time-ordering operation. This observation suggests a straightforward perturbative computation of $\bM$ provided that we identify the unperturbed Hamiltonian with that of the free particle, i.e., $\sH(\tau)=0$. In optical applications this corresponds to $\fn_0=n_0=1$. Performing the first order Born approximation for this system paves the way for a general characterization of perturbative unidirectional reflectionlessness and invisibility
for material with $\fn_0=n_0=1$, \cite{pra-2014a}. This is however overly restrictive, because the condition $\fn_0=n_0=1$ can only be realized for gaseous active media which are known to have very low gain and loss coefficients \cite{silfvast}.

\section{Perturbative Series Expansion for the Transfer Matrix}

Consider the truncated potentials
    \[v_\tau(x):=\left\{\begin{array}{ccc}
    v(x)&{\rm for}& x\leq \tau/k,\\
    0&{\rm for}& x> \tau/k,\end{array}\right.\]
where $\tau$ is a real number, and let $\bI$ stand for the $2\times 2$ unit matrix. Then, the transfer matrix of $v_\tau(x)$, which we denote by $\bM(\tau)$, satisfies \cite{ap-2014,pra-2014a}
    \begin{align}
    &i\partial_\tau \bM(\tau)=\sH(\tau)\bM(\tau),
    \label{M-SE}\\
    & \bM(-\infty)=\bI,~~~~~~~\bM(\infty)=\bM.
    \label{ini}
    \end{align}

Now, suppose that we can express $v(x)$ in the form
    \be
    v(x)=v^{(0)}(x)+v^{(1)}(x),
    \ee
where $v^{(j)}(x)$, with $j=0,1$, are scattering potentials with $v^{(1)}(x)$ playing the role of a perturbation. Furthermore, let
    \begin{align*}
    & w_j(\tau):=\frac{v^{(j)}(\tau/k)}{2k^2}, &&
    \sH_j(\tau):=w_j(\tau)\cK(\tau), &&
    v^{(j)}_\tau(x):=\left\{\begin{array}{ccc}
    v^{(j)}(x)&{\rm for}& x\leq \tau/k,\\
    0&{\rm for}& x> \tau/k,\end{array}\right.
    \end{align*}
$\bM_0$ and $\bM_0(\tau)$ be respectively the transfer matrix of $v^{(0)}(x)$ and $v^{(0)}_\tau(x)$, $\widehat\bM:=\bM_0^{-1}\bM$, and
    \be
    \widehat\bM(\tau):=\bM_0(\tau)^{-1}\bM(\tau).
    \label{tilde-M}
    \ee
Then,
    \begin{align}
    &\sH(\tau)=\sH_0(\tau)+\sH_1(\tau),
    \label{H-H}\\[3pt]
    &i\partial_\tau \bM_0(\tau)=\sH_{0}(\tau)\bM_0(\tau),
    \label{M0-SE}\\[3pt]
    &\bM_0(-\infty)=\bI,~~~~~~~\bM_0(\infty)=\bM_0.
    \end{align}

If we solve (\ref{tilde-M}) for $\bM(\tau)$ and substitute the resulting expression together with (\ref{H-H}) and (\ref{M0-SE}) in (\ref{M-SE}), we find
    \begin{align}
    &i\partial_\tau \widehat\bM(\tau)=\widehat\sH(\tau)\widehat\bM(\tau),
    \label{q1}
    \\
    &\widehat\bM(-\infty)=\bI,~~~~~~~\widehat\bM(\infty)=\widehat\bM,
    \label{q2}
    \end{align}
where
    \bea
    \widehat\sH(\tau)&:=&\bM_0(\tau)^{-1}\sH_{1}(\tau)\,\bM_0(\tau)=w_1(\tau)\widehat\cK(\tau),
    \label{q3}\\
    \widehat\cK(\tau)&:=&\bM_0(\tau)^{-1}\cK(\tau)\,\bM_0(\tau).
    \label{q13}
    \eea
We can express (\ref{q1}) and (\ref{q2}) in the form
    \be
    \widehat\bM=\sT\exp\left\{-i\int_{-\infty}^\infty d\tau' \widehat\sH(\tau')\right\} =\bI+\sum_{\ell=1}^\infty\widehat\bM^{(\ell)},
    \label{q11}
    \ee
where
    \bea
    \widehat\bM^{(\ell)}&:=&(-i)^\ell\int_{-\infty}^\infty d\tau_\ell \int_{-\infty}^{\tau_\ell} d\tau_{\ell-1}\cdots
    \int_{-\infty}^{\tau_2} d\tau_1 \widehat\sH (\tau_\ell) \widehat\sH(\tau_{\ell-1})\cdots\widehat\sH(\tau_1)\nn\\
    &=&(-i)^\ell\int_{-\infty}^\infty d\tau_\ell \int_{-\infty}^{\tau_\ell} d\tau_{\ell-1}\cdots
    \int_{-\infty}^{\tau_2} d\tau_1\widehat\cK(\tau_\ell)\widehat\cK(\tau_{\ell-1})\cdots\widehat\cK(\tau_1)\prod_{p=1}^\ell w_1(\tau_p)\nn\\
    &=&\frac{1}{(2ik)^\ell}\int_{-\infty}^\infty dx_\ell \int_{-\infty}^{x_\ell} dx_{\ell-1}\cdots
    \int_{-\infty}^{x_2} dx_1\widehat\cK(k x_\ell) \widehat\cK(k x_{\ell-1})\cdots
    \widehat\cK(k x_1)\prod_{p=1}^\ell v^{(1)}(x_p).
    \label{series}
    \eea
Note also that
    \be
    \bM=\bM_0\widehat\bM=\sum_{n=0}^\infty\bM^{(n)},
    \label{q5}
    \ee
where
    \begin{align}
    &\bM^{(n)}:=\left\{\begin{array}{ccc}
    \bM_0 & {\rm for} & n=0,\\[3pt]
    \bM_0\widehat\bM^{(n)} &{\rm for}&n\geq 1.\end{array}\right.
    \label{q21}
	\end{align}
Truncating the series on the right-hand side of (\ref{q5}), we obtain approximate perturbative expressions for the transfer matrix of the form
	\be
	\bM\approx \sum_{n=0}^N\bM^{(n)},
	\label{N-order}
	\ee
where $N$ is the order of the approximation (perturbation).

\section{Unidirectional Invisibility in Modulated Refractive Index Profiles}

Consider the optical potentials of the form (\ref{v=21}). Then, for $x\notin[0,L]$,  $v^{(0)}(x)=v^{(1)}(x)=0$, while for $x\in[0,L]$,
   \begin{align}
   & v^{(0)}(x)\approx v_0=k^2(1-\fn_{\!0}^2), &&
   v^{(1)}(x)\approx v_1(x)=-2k^2\fn_{\!0}[\nu(x)+i\kappa(x)].
   \label{v1=k-}
   \end{align}
In particular $v_\tau^{(0)}(x)$ is a barrier potential with a constant hight whose transfer matrix $\bM_0(\tau)$ can be calculated in a closed and exact form \cite{pra-2013a}. For $\tau\in[0,kL]$, it reads
    \be
    \bM_0(\tau)=\left[
    \begin{array}{cc}
    \left[\cos(\fn_{\!0}\tau)+i\fn_+\sin(\fn_{\!0}\tau)\right]e^{-i\tau} & i\fn_- \sin(\fn_{\!0}\tau)\,e^{-i\tau}\\[6pt]
    -i\fn_-\sin(\fn_{\!0}\tau)\,e^{i\tau} & \left[\cos(\fn_{\!0}\tau)-i\fn_+\sin(\fn_{\!0}\tau)\right]e^{i\tau}
    \end{array}\right],
    \label{M0=}
    \ee
where $\fn_\pm:=(\fn_{\!0}\pm\fn_{\!0}^{-1})/2$. We also have
	\be
	\bM_0(\tau)=\left\{\begin{array}{ccc}
	\bI & {\rm for} & \tau\leq 0,\\
	\bM_0(k L) & {\rm for} & \tau\leq k L.\end{array}\right.
	\label{M0=I-M}
	\ee
In particular,
	\be
	\bM^{(0)}=\bM_0=\bM_0(k L).
	\label{M0=2}
	\ee

In view of  (\ref{series}) and the fact that $v^{(1)}(x)$ vanishes for $x\notin[0,L]$, the calculation of $\widehat\bM^{(\ell)}$ and consequently $\widehat\bM$ and $\bM$ only requires the evaluation of $\widehat\cK(\tau)$ for $\tau\in[0,kL]$. To do this we substitute (\ref{K=}) and (\ref{M0=}) in (\ref{q13}). Simplifying the resulting equation we then find
    \be
    \widehat\cK(\tau)=\left[
    \begin{array}{cc}
    \cos^2(\fn_{\!0}\tau)+\fn_{\!0}^{-2}\sin^2(\fn_{\!0}\tau) &
    \left[\cos(\fn_{\!0}\tau)-i\fn_{\!0}^{-1}\sin(\fn_{\!0}\tau)\right]^2\\[6pt]
    -\left[\cos(\fn_{\!0}\tau)+i\fn_{\!0}^{-1}\sin(\fn_{\!0}\tau)\right]^2
    & -\cos^2(\fn_{\!0}\tau)-\fn_{\!0}^{-2}\sin^2(\fn_{\!0}\tau)
    \end{array}\right],
    \label{tK=}
    \ee
where $\tau\in[0,kL]$. Again, because  $v^{(1)}(x)=0$ for $x\notin[0,L]$, as far as (\ref{series}) is concerned we can treat (\ref{tK=}) as if it holds for all $\tau\in\R$.

Inserting (\ref{tK=}) in (\ref{series}), expressing $\cos(\fn_{\!0}kx)$ and $\sin(\fn_{\!0}kx)$ in terms of $e^{\pm i \fn_{\!0} k x}$, and using the expression for the Fourier transform of a function $u(x)$, namely,
	\be
    \tilde u(k):=\int_{-\infty}^\infty dx \, e^{-ikx}u(x),
    \label{Fourier}
    \ee
we can determine $\widehat\bM^{(1)}$. Substituting the result in (\ref{q21}), employing (\ref{M0=I-M}) and (\ref{M0=}), and introducing
    \begin{align}
    &\tilde v_\pm(k):=\tilde v^{(1)}(\pm 2\fn_{\!0}k), &&
    \tilde v_0:=\tilde v^{(1)}(0),
    \label{vv=vv}
    \end{align}
we find the following expressions for the entries of $\bM^{(1)}$.
    \bea
	M^{(1)}_{11}(k)=M^{(1)}_{22}(-k)&=&\frac{e^{-ikL}}{8ik\fn_{\!0}^2}\Big\{
    (\fn_{\!0}^2-1)\left[e^{i\fn_{\!0} kL}\,\tilde v_+(k)+e^{-i\fn_{\!0} kL}\,\tilde v_-(k)\right]\nn\\
    &&\hspace{13mm}+\left[2(\fn_{\!0}^2+1)\cos(\fn_{\!0} kL)+4i\fn_{\!0}\sin(\fn_{\!0} kL)\right]\tilde v_0\Big\},
	\label{M1-11=}\\
    M^{(1)}_{12}(k)=M^{(1)}_{21}(-k)&=&\frac{e^{-ikL}}{8ik\fn_{\!0}^2}\Big\{
    (\fn_{\!0}+1)^2 e^{i\fn_{\!0} kL}\,\tilde v_+(k)+(\fn_{\!0}-1)^2e^{-i\fn_{\!0} kL}\,\tilde v_-(k)\nn\\
    &&\hspace{13mm}+2(\fn_{\!0}^2-1)\cos(\fn_{\!0} kL)\,\tilde v_0\Big\}.
	\label{M1-12=}
   	\eea
We can use these relations together with (\ref{N-order}) and (\ref{M0=}) to give a first-order treatment of the finite-range perturbations of a real or complex rectangular barrier potential. As we alluded to above, this should provide accurate results in optical applications, where $v^{(1)}(x)$ is given by (\ref{v1=k-}). We must however note that due to the $k$-dependence of this potential, we can use (\ref{M1-11=}) and (\ref{M1-12=}) provided that we set
    \begin{align}
    &\tilde v_\pm(k):= -2k^2\fn_{\!0}[\tilde\nu(\pm 2\fn_{\!0} k)+i\tilde\kappa(\pm 2\fn_{\!0} k)],
    &&\tilde v_0:=-2k^2\fn_{\!0}[\tilde\nu(0)+i\tilde\kappa(0)].
    \label{tilde-v1=k-}
    \end{align}

Suppose that we wish to characterize unidirectionally or bidirectionally reflectionless configurations of a generic optically active slab using our first order perturbative scheme. Without loss of generality, we can take $n_0'=0$ so that $\fn_0=n_0\geq 1$, i.e., we consider the barrier potential $v(x)$ of the form (\ref{v=21}) with
    \be
    v_0=k^2(1-n_0^2).
    \label{v0-optics}
    \ee
This is unidirectionally reflectionless to the first order of perturbation theory, if the unperturbed barrier potential is reflectionless to the zeroth order of perturbation theory. In view of (\ref{M0=}) and (\ref{M0=2}), this happens whenever ${n_0} k L$ differs from an integer multiple of $\pi$ by a term $k_1$ that is of order one or higher in the perturbation parameter;
	\be
    k=k_0+k_1,~~~~~~k_0:=\frac{\pi m_0}{{n_0} L},~~~~~m_0=1,2,3,\cdots,~~~~~|k_1|\ll k_0.
	\label{kL=m1}
	\ee
We can use this relation to simplify the expression for the entries of $\bM^{(0)}$ and $\bM^{(1)}$. This gives
    \begin{align}
    &M^{(0)}_{11}\approx e^{-i\mu}(1+iX_+)\approx M^{(0)*}_{22}, && M^{(0)}_{12}
    \approx ie^{-i\mu}X_-\approx M^{(0)*}_{21},
    \label{M-zero=refless}\\
    &M^{(1)}_{11}\approx e^{-i\mu}Y_0,
    &&  M^{(1)}_{12}\approx e^{-i\mu}Y_+,
    \label{M1-11-exp-3}\\
    &M^{(1)}_{21}\approx -e^{i\mu}Y_-,
    && M^{(1)}_{22}\approx -e^{i\mu}Y_0,
    \label{M1-22-exp-3}
    \end{align}
where we have employed (\ref{M0=}), (\ref{M0=2}), (\ref{M1-11=}), (\ref{M1-12=}), and (\ref{kL=m1}), introduced
    \bea
    \mu&:=&\pi m_0(1+n_0^{-1}),~~~~~~~~~~~~~~X_\pm:=\frac{1}{2}(n_0^2\pm 1)k_1L,
    \label{mu=}\\
    Y_0&:=& \frac{(n_0^2-1)[\tilde v_+(k_0)+\tilde v_-(k_0)]+2(n_0^2+1)\tilde v_0}{8ik_0n_0^2}
    \label{X-zer0},\\
    Y_\pm&:=& \frac{(n_0+1)^2\tilde v_\pm(k_0)+(n_0-1)^2\tilde v_\mp(k_0)+2(n_0^2-1)\tilde v_0}{8ik_0n_0^2},
    \label{X-pm}
    \eea
and used `$\approx$' to mean that we ignore quadratic and higher order terms in powers of $k_1$.

The potential $v(x)$ is invisible from the left or right to the first order of perturbation theory, if the unperturbed barrier potential has the same property to the zeroth order of perturbation theory. According to (\ref{M-zero=refless}) this holds if and only if $e^{i\mu}=1$, alternatively $n_0$ is a rational number of the form
    \be
    n_0=\frac{m_0}{2j_0-m_0},~~~~~~~~~j_0=1,2,3,\cdots.
    \label{n-rational}
    \ee
Note that because $n_0\geq 1$, $j_0$ must satisfy $m_0< 2j_0\leq 2m_0$.

The following are simple consequences of (\ref{M-zero=refless}) -- (\ref{X-pm}).
    \begin{itemize}
    \item  $R^l\approx 0$ , if in addition to (\ref{kL=m1}) we have $M^{(0)}_{21}+M^{(1)}_{21}\approx 0$. The latter  means
        \be
        (n_0+1)^2\tilde v_-(k_0)+(n_0-1)^2\tilde v_+(k_0)+2(n_0^2-1)(\tilde v_0-
        2n_0^2k_0k_1L)=0.
        \label{RL=0}
        \ee
    \item  $R^r\approx 0$ , if in addition to (\ref{kL=m1}) we have $M^{(0)}_{12}+M^{(1)}_{12}\approx 0$. This is equivalent to
        \be
        (n_0+1)^2\tilde v_+(k_0)+(n_0-1)^2\tilde v_-(k_0)+2(n_0^2-1)(\tilde v_0-
        2n_0^2k_0k_1L)=0.
        \label{RR=0}
        \ee
    \item $T\approx 1$, if (\ref{kL=m1}) holds together with $M^{(0)}_{22}+M^{(1)}_{22}\approx 1$. This is the case whenever
        \be
        (n_0^2-1)[\tilde v_+(k_0)+\tilde v_-(k_0)]+2(n_0^2+1)(\tilde v_0-
        2n_0^2k_0k_1L)=0.
        \label{T=1}
        \ee
    \end{itemize}

Now, we are in a position to examine the conditions for the perturbative invisibility of the potential $v(x)$. This potential is perturbatively left-invisible provided that (\ref{kL=m1}), (\ref{n-rational}), (\ref{RL=0}) and (\ref{T=1}) hold whereas (\ref{RR=0}) is violated. For $n_0> 1$, which is the case of our interest, we can write (\ref{RL=0}) and (\ref{T=1}) as
    \be
    \tilde v_\pm(k_0)+\left(\frac{n_0\pm 1}{n_0\mp 1}\right)(\tilde v_0-2n_0^2k_0k_1L)=0.
    \label{L-inv-condi}
    \ee
Similarly, $v(x)$ is right-invisible if and only if, in addition to (\ref{kL=m1}) and (\ref{n-rational}), (\ref{RR=0}) and (\ref{T=1}) are satisfied but (\ref{RL=0}) is violated. For $n_0> 1$, we can express (\ref{RR=0}) and (\ref{T=1}) in the form
    \be
    \tilde v_\pm(k_0)+\left(\frac{n_0\mp 1}{n_0\pm 1}\right)(\tilde v_0-2n_0^2k_0k_1L)=0.
    \label{R-inv-condi}
    \ee

In order for the potential to be perturbatively bidirectionally invisible, for $n_0> 1$, (\ref{L-inv-condi}) and (\ref{R-inv-condi}) must hold simultaneously. This implies
    \begin{align}
    &\tilde v_\pm(k_0)=0, && \tilde v_0=2n_0^2k_0k_1L,
    \label{bi-inv-condi}
    \end{align}
which also apply for the case $n_0=1$. Notice that the second of these relations sets the imaginary part of $\tilde v_0$ to zero and determines $k_1$. In view of (\ref{tilde-v1=k-}) and (\ref{kL=m1}), we can express (\ref{bi-inv-condi}) as
    \begin{align}
    &\tilde \nu(\pm 2\pi m_0/L)+i\tilde \kappa(\pm 2\pi m_0/L)=0,
    &&k_1=-\frac{\pi m_0[\tilde\nu(0)+i\tilde\kappa(0)]}{n_0^2L^2}\in\R.
    \label{bi-inv-condi-5}
    \end{align}

For unidirectionally invisible configurations having $n_0>1$, both of the equations in (\ref{bi-inv-condi}) are violated and we can use (\ref{L-inv-condi}) and (\ref{R-inv-condi}) to show that
    \be
    \frac{\tilde v_+(k_0)}{\tilde v_-(k_0)}=
    \left(\frac{n_0-1}{n_0+1}\right)^{\!\!2 {\displaystyle \epsilon}},~~~~
    \epsilon:=\left\{\begin{array}{ccc}
            -1&{\rm for}&\mbox{left-invisibility},\\
            1&{\rm for}&\mbox{right-invisibility}.\end{array}\right.
    \label{necessary-condi}
    \ee
In particular, we have the following criterion for perturbative unidirectional invisibility.
    \begin{itemize}
    \item[] \textbf{Theorem~1}: Let $v(x)$ be a finite-range potential of the form (\ref{v=21})
    with $\fn_0=n_0>1$. Then a necessary condition for the perturbative unidirectional invisibility of $v(x)$ is that $\tilde v_+(k_0)/\tilde v_-(k_0)$ be given by (\ref{necessary-condi}) for some $k_0$ of the form (\ref{kL=m1}). In particular this quantity must take a real and positive value.
    \end{itemize}
As a simple application of this theorem consider the potential
    \be
    v(x)=\left\{\begin{array}{cc}
    v_0+ \fa\, x+\fb\, x^2 & {\rm for}~x\in[0,L],\\
    0 &{\rm otherwise},\end{array}\right.
    \label{v=a-b}
    \ee
where $\fa$ and $\fb$ are nonzero complex parameters, and suppose that $k_0$ satisfies (\ref{kL=m1}). Then,
    \[\frac{\tilde v_+(k_0)}{\tilde v_-(k_0)}=2\left[1-\pi i m_0 \left(\frac{\fa}{L\fb}+1\right)\right]^{-1}-1,\]
which is real and positive provided that there is a real number $\xi$ not larger than $\pi^{-1}$ such that $\fa=(-1+i\xi)L\fb$. According to Theorem~1 if this condition is violated, the potential (\ref{v=a-b}) is incapable of supporting perturbative unidirectional invisibility.

Next, let us examine the consequences of a constant real shift of the potential on its support, i.e.,
    \be
    v(x)\to w(x):=\left\{\begin{array}{ccc}
    v(x)+\alpha & {\rm for} &x\in[0,L],\\
    0 & {\rm for} &x\notin[0,L],\end{array}\right.~~~~~~\alpha\in\R.
    \label{shift}
    \ee
It is easy to show that for the values of $k_0$ given by (\ref{kL=m1}), this transformation leaves
$\tilde v_\pm(k_0)$ invariant and changes $\tilde v_0$ by a real additive term, namely $\alpha L$, i.e.,
    \[ \tilde v_\pm(k_0)\to \tilde w_\pm(k_0)=\tilde v_\pm(k_0),~~~~~~~~\tilde v_0\to
    \tilde w_0=\tilde v_0+\alpha L.\]
In particular, if we choose $\alpha=2n_0^2k_0k_1$, then Eqs.~(\ref{RL=0}) -- (\ref{bi-inv-condi}) for the transformed potential $w(x)$ have the same form as those of $v(x)$ with $k_1$ set to zero. More generally, we have the following useful result.
    \begin{itemize}
    \item[] \textbf{Theorem~2}: Let $v(x)$ be as in Theorem~1. Then we can tune the value of $k_1$ and hence the wavenumber $k$ at which $v(x)$ is perturbatively unidirectionally or bidirectional invisible by performing a constant shift of its real part according to (\ref{shift}).
    \end{itemize}

\section{Invisible $\cP\cT$-Symmetric Potentials}

Let $\cP$ and $\cT$ denote the space-reflection and time-reversal operators,
    \[\cP\psi(x):=\psi(L-x),~~~~~~~~~ \cT\psi(x):=\psi(x)^*,\]
and consider a $\cP\cT$-symmetric optical potential given by (\ref{v=21}) and (\ref{v=21-def}), that by definition satisfies $v(L-x)^*=v(x)$. Then, without loss of generality, we can take $\fn_0=n_0\in\R$, so that $v_0=k^2(1-n_0^2)$ is real. This in turn implies
    \begin{align}
    &\nu(L-x)=\nu(x), &&\kappa(L-x)=-\kappa(x).
    \end{align}
Using these relations and the fact that $\nu(x)$ and $\kappa(x)$ vanish for $x\notin[0,L]$, we find that
    \begin{align}
    &\tilde\nu(k)=2\,e^{-ikL/2}\int_0^{L/2} dx \cos[k(\mbox{$\frac{L}{2}$}-x)]\,\nu (x),
    \label{Four-nu}\\
    &\tilde\kappa(k)=-2i\,e^{-ikL/2}\int_0^{L/2} dx \sin[k(\mbox{$\frac{L}{2}$}-x)]\,\kappa (x).
    \label{Four-ka}
    \end{align}
In particular, because $n_0k_0=\pi m_0/L$ and $m_0$ is an integer, $\tilde\nu(\pm 2n_0k_0)$ and $\tilde\kappa(\pm 2n_0k_0)$ respectively take real and imaginary values. It is also easy to see that
    \begin{align}
    &\tilde\nu(-2n_0 k_0)=\tilde\nu(2n_0 k_0), &&\tilde\kappa(-2n_0 k_0)=-\tilde\kappa(2n_0 k_0),
    \label{parity=}\\
    &\tilde\nu(0)=\int_0^L dx\:\nu (x), &&\tilde\kappa(0)=0.
    \end{align}

In Ref.~\cite{pra-2013a}, we show that the equations governing the phenomenon of unidirectional invisibility have an intrinsic $\cP\cT$-symmetry. This makes $\cP\cT$-symmetric potentials the primary class of potentials with this property. An interesting manifestation of this observation is the fact that for $\cP\cT$-symmetric potentials the quantity $\tilde v_-(k_0)/\tilde v_+(k_0)$ is always real (See Theorem~1.) This follows from (\ref{tilde-v1=k-}), (\ref{kL=m1}), and (\ref{parity=}) and the above-mentioned reality of  $\tilde\nu(\pm 2n_0k_0)$ and $i\tilde\kappa(\pm 2n_0k_0)$. More  generally, for $\cP\cT$-symmetric potentials, we can express Condition (\ref{necessary-condi}) of Theorem~1 as
    \be
    \tilde\kappa(2n_0 k_0)=\frac{2i\epsilon\, n_0\,\tilde\nu(2n_0k_0)}{n_0^2+1}.
    \label{necessary-condi-2}
    \ee

Next, we consider the cases where $\nu(x)$ is a constant. Then (\ref{Four-nu}) implies $\tilde\nu(2n_0 k_0)=0$. This in turn reduces (\ref{necessary-condi-2}) to $\tilde\kappa(2n_0k_0)=0$ and leads to $\tilde v_\pm(k_0)=0$. As we discussed earlier, this marks the perturbative bidirectional invisibility of $v(x)$ and proves the following theorem.
    \begin{itemize}
    \item[] \textbf{Theorem~3:} Let $v(x)$ be a $\cP\cT$-symmetric potential of the form (\ref{v=21}). Suppose that the real part of $v(x)$ takes a constant value on its support $[0,L]$. Then $v(x)$ cannot display perturbative unidirectional invisibility.
    \end{itemize}
This theorem shows that one cannot realize unidirectional invisibility by engineering the loss-gain profile of an optically active material obtained by doping a homogeneous host medium; the real part of the refractive index must also be modulated properly.

Employing (\ref{parity=}) in (\ref{tilde-v1=k-}), we obtain
    \begin{align}
    &\tilde v_\pm(k_0)=-2n_0k_0^2[\tilde\nu(2n_0k_0)\pm i\tilde\kappa(2n_0k_0)],
    &&\tilde v_0=-2n_0k_0^2\tilde\nu(0).
    \label{parity=2}
    \end{align}
These relations simplify the conditions (\ref{RL=0})  -- (\ref{T=1}) for perturbative reflectionlessness and transparency of the potential $v(x)$ and lead to the following observations.
    \begin{itemize}
    \item $v(x)$ is perturbatively reflectionless from the left $(\epsilon=-1)$ or right $(\epsilon=1)$ at the wavenumber $k=k_0+k_1$ provided that
            \be
            (n_0^2+1)\tilde\nu(2n_0k_0)+2i\epsilon\,n_0\tilde\kappa(2n_0k_0)+
            (n_0^2-1)\left[\tilde\nu(0)+\frac{n_0Lk_1}{k_0}\right]=0.
            \label{PT-RR=0}
            \ee
    \item It is perturbatively transparent at this wavenumber if and only if $n_0$ and $k_1$ are respectively given by (\ref{n-rational}) and
        \be
        k_1=-\frac{k_0}{n_0(n_0^2+1)L}\left[(n_0^2-1)\tilde\nu(2n_0k_0)+(n_0^2+1)\tilde\nu(0)\right].
        \label{perfect-trans-k1}
        \ee
    \end{itemize}
These observations lead to the following characterization of perturbative unidirectional invisibility for $\cP\cT$-symmetric potentials.
    \begin{itemize}
    \item[] \textbf{Theorem~4:} Let $v(x)$ be a $\cP\cT$-symmetric potential given by  (\ref{v=21}) and (\ref{v=21-def}), $j_0$ and $m_0$ be positive integers, $\fn_0=n_0=(2j_0/m_0-1)^{-1}\geq 1$, and $k_0:=\pi m_0/n_0L$. Then $v(x)$ displays perturbative unidirectional invisibility for the wavelength $k=k_0+k_1$ if and only if  (\ref{necessary-condi-2}) and (\ref{perfect-trans-k1}) hold and $\tilde\nu(2n_0k_0)\neq 0$.
    \end{itemize}

Let us also note that Eqs.~(\ref{PT-RR=0}) and (\ref{perfect-trans-k1}) simplify considerably for $n_0=1$. In this case they imply that $v(x)$ is perturbatively unidirectionally reflectionless if
    \be
    \tilde\nu(2k_0)=-i\epsilon\,\tilde\kappa(2k_0)\neq 0,
    \label{PT-RR=0-n}
    \ee
and perturbatively transparent if
    \be
    k_1=-\frac{k_0\tilde\nu(0)}{L}.
    \label{perfect-trans-k1-n}
    \ee
The fact that $k_1$ does not enter in (\ref{PT-RR=0-n}) seems to indicate that perturbative reflectionlessness is not sensitive to small changes of the wavelength whenever $n_0=1$. This is consistent with the known results for specific $\cP\cT$-symmetric potentials considered in the literature (See for example \cite{pra-2014a}.)

Next, we observe that, because $n_0k_0=\pi m_0/L$, we can write (\ref{necessary-condi-2}) as
    \be
    \tilde f_\epsilon(\mbox{\large$\frac{2\pi m_0}{L}$})=0,
    \label{necessary-condi-2-new}
    \ee
where
    \be
    f_\pm(x):=\nu(x)\pm\frac{i(n_0^2+1)}{2n_0}\,\kappa(x).
    \label{f-pm=}
    \ee
This suggests that any $\cP\cT$-symmetric potential (\ref{v=21}) for which the right-hand side of (\ref{f-pm=}) differs from that of $v(x)$ by a constant multiplicative factor will have similar unidirectional invisibility property as $v(x)$. The following is a precise statement of this result.
    \begin{itemize}
    \item[] \textbf{Theorem~5:} Let $v(x)$ and $\check v(x)$ be $\cP\cT$-symmetric potentials of the form (\ref{v=21}) with the corresponding refractive indices $\fn(x)=n_0+\nu(x)+i\kappa(x)$ and $\check \fn(x)=\check n_0+\check\nu(x)+i\check\kappa(x)$, where $n_0=(2j_0/m_0-1)^{-1}\geq 1$, $\check n_0=(2\check j_0/m_0-1)^{-1}\geq 1$, $j_0, \check j_0$, and $m_0$ are positive integers, and $\nu,\kappa,\check\nu$ and $\check\kappa$ are real-valued functions vanishing outside $[0,L]$.  Let $k_0:=\pi m_0/n_0L$ and $k_1$ be given by (\ref{perfect-trans-k1}), and suppose that there is a nonzero real number $\alpha$ of the order of 1 such that
            \begin{align}
            &\check\nu(x)=\alpha\,\nu(x), && \check\kappa(x)=\frac{\alpha\,\check n_0(n_0^2+1)\kappa(x)}{n_0(\check n_0^2+1)}.
            \label{thm5-condi}
            \end{align}
        Then $v(x)$ is perturbatively left-invisible (respectively right-invisible) for the wavenumber $k=k_0+k_1$ if and only if $\check v(x)$ is perturbatively  left-invisible (respectively right-invisible) for the wavenumber $\check k:=n_0k_0/\check n_0+\check k_1$, where
            \[\check k_1:=-\frac{\pi m_0\alpha}{\check n_0^2(\check n_0^2+1)L^2}\left[(\check n_0^2-1)\tilde\nu(\mbox{\large$\frac{2\pi m_0}{L}$})+
            (\check n_0^2+1)\tilde\nu(0)\right].\]
    \end{itemize}

For example, consider the case where $\alpha=\check n_0=1$,
    \begin{align}
    &\check \nu(x)=\nu(x)=\nu_0\,\cos(\mbox{\large$\frac{2\pi m_0 x}{L}$}),
    && \check\kappa(x)=\frac{(n_0^2+1)\kappa(x)}{2n_0} =
    \nu_0\sin(\mbox{\large$\frac{2\pi m_0 x}{L}$}),
    \label{check=}
    \end{align}
$\nu_0\in\R$, and $x\in[0,L]$. Then the hypothesis of Theorem~5 holds for the refractive index profiles
    \bea
     \fn(x)&=& n_0 + \nu_0\left[\cos(\mbox{\large$\frac{2\pi m_0 x}{L}$})+\frac{2 i n_0 }{n_0^2+1}\,\sin(\mbox{\large$\frac{2\pi m_0 x}{L}$})\right],
    \label{v-non-exp}\\
    \check\fn(x)&=&1+\nu_0\, e^{2\pi i m_0 x/L}.
    \label{v-exp-1}
    \eea
Because the latter is perturbatively left-invisible for the wavelengths $\check k=\pi m_0 /L$, \cite{pra-2014a}, according to Theorem~5, the former should be perturbatively left-invisible for some wavelength $k=\pi m_0 /n_0 L+k_1$. Using (\ref{check=}), we can easily show that $\tilde\nu(\mbox{\small $2\pi m_0/L$})=\nu_0 L/2$ and $\tilde\nu(0)=0$. Substituting these relations in (\ref{perfect-trans-k1}) and setting $k_0=\pi m_0/n_0L$ give
    \be
    k_1=-\frac{\pi m_0(n_0^2-1)\nu_0}{2n_0^2(n_0^2+1)L}.
    \label{k1=non-exp}
    \ee

\begin{figure}
	\begin{center}
	\includegraphics[scale=.6]{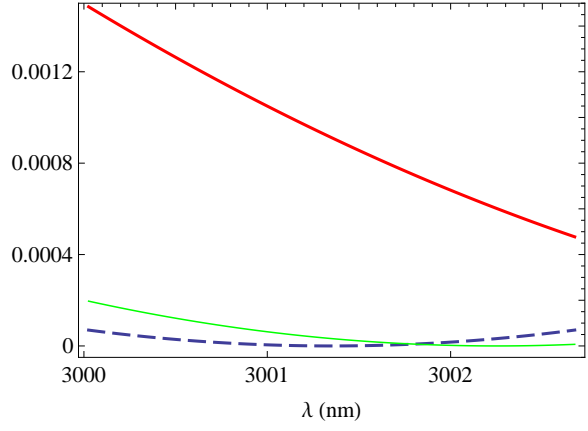}
	\caption{Graphs of $|R^l|^2$ (navy dashed curve), $|R^r|^2$ (thick solid red curve) and $|T-1|^2$ (solid green curve) as functions of the wavelength $\lambda$ for the refractive index profile given by (\ref{v-non-exp})  and (\ref{n-spec1}). The fact that $|R^l|^2$ and $|T-1|^2$ take much smaller values than $|R^r|^2$ is an evidence of the unidirectional invisibility of this index profile.}
	\label{fig1}
	\end{center}
	\end{figure}

We have checked the above predictions by numerically evaluating $R^{l/r}$ and $T$ for the index profile (\ref{v-non-exp}) with the following numerical values for its parameters.
    \begin{align}
    n_0=2 &&\nu_0=3\times 10^{-3}, &&L=6~\mu{\rm m}, && m_0=8.
    \label{n-spec1}
    \end{align}
This choice of $m_0$ yields $n_0=2$ for $j_0=6$ and $\check n=1$ for $\check j_0=8$. In view of (\ref{k1=non-exp}) and (\ref{n-spec1}), $k_1=-9.424778\times 10^{-4}/\mu{\rm  m}$. We also find for the wavenumber and the wavelength at which (\ref{v-non-exp}) is perturbatively left-invisible, $k=2.093452/\mu{\rm  m}$ and  $\lambda=3001.35~{\rm  nm}$, respectively. Figure~\ref{fig1} shows the plots of the reflection coefficients $|R^{l/r}|^2$ and the quantity $|T-1|^2$ confirming the validity of our approximate (perturbative) results concerning left-invisibility of the index profile (\ref{v-non-exp}) for the parameter values (\ref{n-spec1}).
Let us also note that for these values, the index profile (\ref{v-exp-1}) is left-invisible for the wavelength $\check \lambda=2L/m_0=1500~{\rm nm}$.

\section{Invisible Locally Periodic Perturbations and Their Superpositions}

Consider the locally periodic potential
    \be
    v(x)=\left\{\begin{array}{cc}
    v_0+\fz\, e^{iK x}&{\rm for}~x\in[0,L],\\[6pt]
    0 & {\rm otherwise},\end{array}\right.
    \label{v-lp1}
    \ee
where $v_0$, $\fz$, and $K$ are real parameters, and $\fz\neq 0\neq K$. In optical applications, $v_0$ is given by (\ref{v0-optics}) and
	\be
	\fz=-2k^2n_0\nu_0,	
	\label{optics=}
	\ee
where $\nu_0$ is a real parameter determining the complex refractive index of the medium according to (\ref{n2=x}) with
	\begin{align}
	& \fn_0=n_0, &&\nu(x)=\nu_0\cos(K x), && \kappa(x)=\nu_0\sin(K x).
	\end{align}
For $n_0=1$, we have $v_0=0$, and (\ref{v-lp1}) with $K$ being an integer multiple of $2\pi/L$ is the primary example of a unidirectionally invisibility potential considered in the literature \cite{invisible-0,lin,invisible-1,pra-2014a}. In what follows we explore the invisibility properties of the potential (\ref{v-lp1}) for arbitrary $n_0\geq 1$ and $K\in\R$.

For the potential (\ref{v-lp1}), $v^{(1)}(x)=\fz\, e^{iK x}$ and (\ref{Fourier}) gives
    \[\tilde v^{(1)}(k)=\frac{i\fz\big[1-e^{i(K-k)L}\big]}{K-k}.\]
Substituting this equation in (\ref{vv=vv}), we obtain $\tilde v_\pm(k)$ and $\tilde v_0$. These together with (\ref{RL=0}) -- (\ref{bi-inv-condi}) allow us to determine the reflectionless and invisible configurations of (\ref{v-lp1}). We describe these by considering the following two cases separately.
\begin{itemize}
    \item[]\textbf{Case I.} $K=\pm 2\pi m/L$ for a positive integer $m$: Then in order for the potential to have perturbative unidirectional reflectionlessness or perfect transmission, we must have $m=m_0$ so that
            \be
            K=\pm 2n_0k_0=\pm\frac{2\pi m_0}{L}.
            \label{K-integer}
            \ee
    Furthermore, the following hold.
    \begin{itemize}
    \item[]I.1) The potential (\ref{v-lp1}) is perturbatively left- or right-reflectionless whenever
            \begin{align}
            &k_1=\frac{(n_0\pm\epsilon)\fz}{4n_0^2(n_0\mp\epsilon)k_0}=
            \frac{-(n_0\pm\epsilon)k_0\nu_0}{2n_0(n_0\mp\epsilon)}.
            \end{align}
        where
            \be
            \epsilon:=\left\{\begin{array}{ccc}
            -1&{\rm for}&\mbox{left-reflectionlessness},\\
            1&{\rm for}&\mbox{right-reflectionlessness}.
            \end{array}\right.
            \label{epsilon-def}
            \ee
    \item[]I.2) It displays perturbative perfect transmission provided that $n_0$ satisfies (\ref{n-rational}) and
            \begin{align}
            &k_1=\frac{(n_0^2-1)\fz}{4n_0^4(n_0^2+1)k_0}=
            \frac{-(n_0^2-1)k_0\nu_0}{2n_0(n_0^2+1)}.
            \end{align}
    \item[]I.3) It supports perturbative unidirectional invisibility if and only if $n_0=1$ and $k_1=0$. This is the case studied in Refs.~\cite{invisible-0,lin,invisible-1,pra-2014a}.
    \end{itemize}

    \item[]\textbf{Case II.} $K$ is not an integer multiple of $2\pi/L$: Then (\ref{v-lp1}) supports unidirectional reflectionlessness or perfect transmission only if $k_1=0$.
        Furthermore, we can establish the following.
    \begin{itemize}
    \item[]II.1) The potential (\ref{v-lp1}) is perturbatively left- or right-reflectionless whenever
            \be
            K=\left(-\epsilon\pm\sqrt{2 n_0^2-1}\right)k_0,
            \ee
        where $\epsilon$ is defined by (\ref{epsilon-def}).
    \item[]II.2) It displays perturbative perfect transmission provided that $n_0$ satisfies (\ref{n-rational}), and
            \be
            K=\pm\sqrt{2(n_0^2+1)}\,k_0.
            \ee
    \item[]II.3) It does not support perturbative unidirectional invisibility for any value of $n_0\geq 1$.
    \end{itemize}

\end{itemize}

The fact that for $n_0>1$ the potential (\ref{v-lp1}) cannot support perturbative unidirectional invisibility motivates the search for its generalizations that possess this property. For example consider the potentials given by:
    \be
    v(x)=\left\{\begin{array}{cc}
    v_0+\fz_0+\displaystyle \sum_{\ell=1}^N\fz_\ell e^{iK_\ell x}&{\rm for}~x\in[0,L],\\[12pt]
    0 & {\rm otherwise},\end{array}\right.
    \label{v-lp1-gen}
    \ee
where $v_0$ is given by (\ref{v0-optics}) for some $n_0\geq 1$, $N\leq\infty$, $\fz_\ell$ are real or complex, $K_\ell$ are real, and $\fz_\ell\neq 0\neq K_\ell$. Then, in view of (\ref{kL=m1}), we find
    \begin{align}
    &\tilde v_\pm(k_0)=\sum_{\ell=1}^N\frac{i\fz_\ell\left(1-e^{iK_\ell L}\right)}{K_\ell\mp 2n_0k_0},
    &&\tilde v_0=\fz_0L+\sum_{\ell=1}^N\frac{i\fz_\ell\left(1-e^{iK_\ell L}\right)}{K_\ell}.
    \label{sum-tv}
    \end{align}

Next, we substitute (\ref{sum-tv}) in (\ref{L-inv-condi}) and (\ref{R-inv-condi}) to determine the conditions for the perturbative unidirectional invisibility of the potential (\ref{v-lp1-gen}). This results in the following pair of necessary conditions for the perturbative left-invisibility.
    \be
    \sum_{\ell=1}^N\fz_\ell F_\ell G^\pm_\ell(k_0)=\frac{i\tilde\fz_0L(n_0\pm 1)}{2n_0},
    \label{L-inv-condi-2}
    \ee
where
    \begin{align}
    &F_\ell:=\frac{1-e^{iK_\ell L}}{K_\ell},
    &&G_\ell^\pm(k):=\frac{K_\ell-(1\pm n_0)k}{K_\ell\mp 2 n_0 k},
    &&\tilde\fz_0:=\fz_0-2n_0^2k_0k_1.
    \label{F-G=}
    \end{align}
Similarly, we find the following equations for perturbative right-invisibility.
    \be
    \sum_{\ell=1}^N\fz_\ell F_\ell G^\pm_\ell(-k_0)=\frac{i\tilde\fz_0L(n_0\pm 1)}{2n_0}.
    \label{R-inv-condi-2}
    \ee
The potential (\ref{v-lp1-gen}) is perturbatively left-invisible (respectively right-invisible) if and only if (\ref{L-inv-condi-2}) (respectively (\ref{R-inv-condi-2})), (\ref{kL=m1}) and (\ref{n-rational}) are satisfied while $\tilde v_0\neq 0$.

For instance consider the case that $\fz_0=0$, $N=2$, and $K_1\neq K_2$, i.e.,
    \be
    v(x)=\left\{\begin{array}{cc}
    v_0+\fz_1 e^{iK_1x} +\fz_2 e^{iK_2x} &{\rm for}~x\in[0,L],\\
    0 & {\rm otherwise}.\end{array}\right.
    \label{v-unidir}
    \ee
Then we can set $k_1=0$, so that $\tilde\fz_0=0$ and the left-invisibility conditions (\ref{L-inv-condi-2}) reduce to a pair of homogeneous linear equations for $\fz_1$ and $\fz_2$. These have a nontrivial solution of the form
    \be
    \fz_2=-\frac{F_1G_1^+(k_0)\,\fz_1}{F_2G_2^+(k_0)},
    \label{zeta2=}
    \ee
provided that
    \bea
    &F_1\neq 0 \neq F_2,&
    \label{F-not-zero}\\
    &G_1^+(k_0)G_2^-(k_0)-G_1^-(k_0)G_2^+(k_0)=0.&
    \label{nontrivial}
    \eea

For $K_1=2k_0$, Eq.~(\ref{nontrivial}) implies $K_2=2k_0$, which violates the condition $K_1\neq K_2$. Therefore, we take $K_1\neq 2k_0$. In view of this relation and (\ref{F-G=}), we can reduce (\ref{nontrivial}) to a quadratic equation for $K_2$ with a pair of solutions, namely $K_2=K_1$, which is inadmissible, and
    \be
    K_2=2k_0\left[1-\frac{(n_0^2-1)k}{K_1-2k_0}\right].
    \label{K2=}
    \ee
We can express this relation in the following more symmetric form \footnote{This is a manifestation of the fact that under the transformation $K_1\leftrightarrow K_2$ is a consequence of the (\ref{nontrivial}) is left invariant under the exchange of $K_1$ and $K_2$.}
    \be
    (K_1-2k_0)(K_2-2k_0)=-2(n_0^2-1)k_0^2,
    \label{K1-K2}
    \ee
which, in particular, implies $K_2\neq 2k_0$.

Next, we examine the consequences of (\ref{F-not-zero}). According to (\ref{F-G=}), this relation implies that either $K_1=-K_2=\pm 2 n_0 k_0=\pm 2\pi m_0/L$, or $K_1$ and $K_2$ are not integer multiples of $2\pi/L$. The first of these possibilities is in conflict with (\ref{K1-K2}). Hence the second must hold.

Substituting (\ref{K2=}) in (\ref{zeta2=}) and using (\ref{kL=m1}) to simplify the result, we find
    \be
    \fz_2=-\frac{4k[K_1-(n_0^2+1)k_0][(K_1-k_0)^2-n_0^2k_0^2](e^{i K_1 L}-1)\,\fz_1}{
    K_1(K_1-2k_0)(K_1^2-4n_0^2k_0^2)(e^{-2ik^2L(n_0^2-1)/(K_1-2k_0)}-1)}.
    \label{z2=}
    \ee

A similar analysis of the necessary conditions for the realization of perturbative right-invisibility of (\ref{v-unidir}) yields:
    \begin{align}
    &K_2=-2k_0\left[1+\frac{(n_0^2-1)k_0}{K_1+2k_0}\right],
    \label{K2-R}\\
    &\fz_2=\frac{4k[K_1+(n_0^2+1)k_0][(K_1+k_0)^2-n_0^2k_0^2](e^{i K_1 L}-1)\,\fz_1}{
    K_1(K_1+2k_0)(K_1^2-4n_0^2k_0^2)(e^{-2ik^2L(n_0^2-1)/(K_1+2k_0)}-1)}.
    \label{z2-R}
    \end{align}
These can respectively be obtained from (\ref{K2=}) and (\ref{z2=}) by taking $k_0$ to $-k_0$. Note also that in (\ref{K2=}) and (\ref{z2-R}), $k_0$ and $n_0$ are given by (\ref{kL=m1}) and (\ref{n-rational}).

Equations (\ref{K2=}) -- (\ref{z2=}) and (\ref{K2-R}) -- (\ref{z2-R}) describe perturbative unidirectionally invisible configurations provided that they do not hold simultaneously. If they do, we obtain a bidirectionally invisible configuration. This happens whenever
    \begin{align}
    &K_1=-K_2=\pm\sqrt{2(n_0^2+1)}\,k_0,
    &&\fz_2=-\fz_1 e^{iK_1L}.
    \label{bidir-condi}
    \end{align}
Substituting these relations in (\ref{v-unidir}) and introducing $\fz:=2 e^{iK_1L/2}\fz_1$, we find
    \be
    v(x)=\left\{\begin{array}{cc}
    v_0+i \fz\sin\left[K_1(x-L/2)\right]&{\rm for}~x\in[0,L],\\
    0 & {\rm otherwise},\end{array}\right.
    \label{v-bidir}
    \ee
which is $\cP\cT$-symmetric for real values of $\fz$.This potential displays perturbative  bidirectional invisibility at the wavenumbers $k=k_0$ provided that $v_0$, $n_0$, and $K_1$ satisfy (\ref{v0-optics}), (\ref{n-rational}) and (\ref{bidir-condi}). If $K_2$ and $\fz_2$ are given by (\ref{K2=}) and (\ref{z2=}) (respectively (\ref{K2-R}) and (\ref{z2-R})) and we use a value of $K_1$ which violates the first equation in (\ref{bidir-condi}), then the potential (\ref{v-unidir}) displays perturbative left- (respectively right-) invisibility.

In the remainder of this section we consider concrete optical implementations of our results for a sample with $n_0=3.4$ which in light of (\ref{v0-optics}) implies
    \be
    v_0=-10.560\,k^2.
    \label{v0=sample}
    \ee
Suppose that we wish to realize perturbative invisibility for $k=k_0=4\pi/3\,\mu{\rm m}$ which corresponds to the wavelength $\lambda=1500\,{\rm nm}$. Then, according to (\ref{n-rational}), we can take $m_0=51$ and $j_0=33$ which together with (\ref{kL=m1}) determine the thickness of the slab to be
    \be
    L=11.250\,\mu{\rm m}.
    \label{thickness}
    \ee

\begin{figure}
	\begin{center}
	\includegraphics[scale=.6]{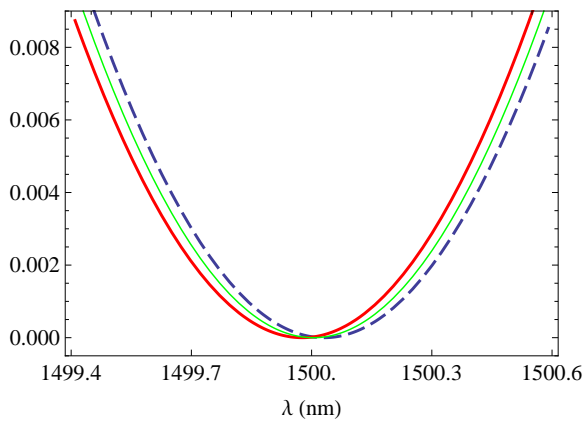}
	\caption{Graphs of $|R^l|^2$ (navy dashed curve), $|R^r|^2$ (thick solid red curve) and $|T-1|^2$ (solid green curve) as functions of the wavelength $\lambda$ for the potential given by (\ref{v-bidir}) -- (\ref{specifics}). The fact that these curves take a zero value for their common minimum at $\lambda=1500~{\rm nm}$ is a clear manifestation of the bidirectional invisibility of the potential.}
	\label{fig2}
	\end{center}
	\end{figure}

Setting $n_0=3.4$ and $k_0=4\pi/3\,\mu{\rm m}$ in (\ref{bidir-condi}), we find that the potential (\ref{v-bidir}) is bidirectionally invisible for this wavelength provided that we take $ K_1=\pm 20.994/\mu{\rm m}$ and $|\fz_1/k_0^2|\ll 1$. Figure~\ref{fig2} gives a graphical demonstration of a direct numerical calculation of $|R^{l/r}|^2$ and $|T-1|^2$ for the potential (\ref{v-bidir}) with
    \begin{align}
    &\fz =0.05\,k_0^2, && K_1=20.994/\mu{\rm m}.
    \label{specifics}
    \end{align}
As seen from this figure, $|R^{l/r}|^2$ and $|T-1|^2$ take very small values for $\lambda=1500\,{\rm nm}$. More specifically, our numerical calculations give $|R^{l/r}|^2< 10^{-5}$ and $|T-1|^2<10^{-7}$. This provides an independent confirmation of our result pertaining perturbative bidirectional invisibility of this potential.

\begin{figure}
	\begin{center}
	\includegraphics[scale=.6]{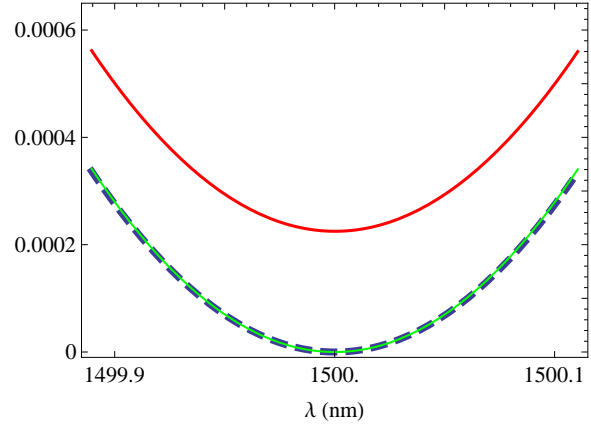}
	\caption{Graphs of $|R^l|^2$ (navy dashed curve), $|R^r|^2$ (thick solid red curve) and $|T-1|^2$ (solid green curve) as functions of the wavelength $\lambda$ for the potential given by (\ref{v-unidir}), (\ref{v0=sample}), (\ref{thickness}), (\ref{specifics2}), and (\ref{specifics3}). The graphs for $|R^l|^2$ and $|T-1|^2$ overlap in the scale depicted here.}
	\label{fig3}
	\end{center}
	\end{figure}

Next, we examine a left-invisible configuration of the potential (\ref{v-unidir}). This requires determining the values of $\fz_2$ and $K_2$ using (\ref{K2=}) and (\ref{z2=}), but does not restrict the choice of $\fz_1$ and $K_1$ except for the fact that $|\fz_1/k^2|\ll 1$, $K_1\neq 8\pi/3\,\mu{\rm m}= 8.37758/\mu{\rm m}$, and $K_1\neq \pm 20.994/\mu{\rm m}$,. We choose
    \begin{align}
    \fz_1=0.08\, k_0^2, && K_1=-17.593/\mu{\rm m},
    \label{specifics2}
    \end{align}
which together with (\ref{K2=}) and (\ref{z2=}) give
    \begin{align}
    \fz_2=(-0.111478 + 0.0170778\,i) k_0^2, && K_2=22.647/\mu{\rm m}.
    \label{specifics3}
    \end{align}
Figure~\ref{fig3} shows the plots of $|R^{l/r}|^2$ and $|T-1|^2$ for the potential (\ref{v-unidir}) with $v_0, L, \fz_1,K_1,\fz_2$, and $K_2$ given by (\ref{v0=sample}), (\ref{thickness}), (\ref{specifics2}), and (\ref{specifics3}).
It clearly confirms the left-invisibility of this potential for $\lambda=1500~{\rm nm}$. Our numerical calculations show that for this wavelength, $|R^l|^2< 10^{-9}$, $|T-1|^2< 10^{-10}$, and $|R^r/R^l|^2> 10^5$.

\section{Concluding Remarks}

In this article we have developed a general perturbative scheme for the study of scattering properties of optical material obtained by modulating a general homogenous medium whose refractive index $n_0$
may exceed unity substantially. We have modeled this problem in terms of a perturbed rectangular barrier potential and conducted a detailed investigation of its scattering features paying particular attention to unidirectionally invisible configurations. The result is a set of basic theorems revealing the general properties of perturbative unidirectional invisibility.

$\cP\cT$-symmetric potentials have a distinctive place in the study of the phenomenon of unidirectional invisibility, for the very equations that define this phenomenon are $\cP\cT$-invariant \cite{pra-2013a}. This leads to a variety of simplifications when one searches for  $\cP\cT$-symmetric unidirectionally invisible potentials. Among the most important results of our investigation is the observation that a $\cP\cT$-symmetric refractive index profile with a constant real part (on its support) cannot display unidirectional invisibility. Another remarkable result is the existence of families of unidirectionally invisible $\cP\cT$-symmetric index profiles whose members are obtained from a seed member by performing certain scaling transformations. These map index profiles with $n_0=1$ to those with $n_0>1$.

We have also examined the locally periodic complex exponential potentials of the form (\ref{v-lp1}) and showed that they support perturbative unidirectional invisibility only for the known $\cP\cT$-symmetric case where $n_0=1$. The superpositions of a finite number of such potentials are, however, capable of displaying this feature even for $n_0>1$. We have constructed specific examples of such superposed locally periodic potentials.

As a final note, we wish to stress that our results lifts a serious limitation on the practical implementation of unidirectional invisibility in optical settings, because it allows for the use of high-gain optical material, which have $n_0>1$, to develop unidirectionally invisible devices.

\vspace{6pt}

\noindent{\em Acknowledgments:}  I would like to thank Ali Serpeng\"{u}zel for bringing to my attention one of the references, and Sasan Haji-Zadeh for carefully reading the first draft of the paper and helping me find and correct a few typos. This work has been supported by  the Scientific and Technological Research Council of Turkey (T\"UB\.{I}TAK) in the framework of the project no: 112T951, and by the Turkish Academy of Sciences (T\"UBA).

\ed